\documentclass[aps,twocolumn,superscriptaddress,nofootinbib]{revtex4-1}

\usepackage[colorlinks,citecolor=blue]{hyperref}
\usepackage{enumerate}
\usepackage{xcolor,graphicx}
\usepackage{latexsym,cancel,amssymb,amsmath,verbatim,mathrsfs}
\usepackage{mathtools}
\usepackage{physics,ulem}

\usepackage{multirow}

\newcommand{\bea}{\begin{eqnarray}}
\newcommand{\beq}{\begin{equation}}

\newcommand{\eea}{\end{eqnarray}}
\newcommand{\eeq}{\end{equation}}

\newcommand{\ld}{\lambda}

\newcommand{\nn}{\nonumber}

\newcommand{\sg}{\sigma}

\newcommand{\lr}{ \xrightleftharpoons[\scriptsize{\text{Constraint}}]{\scriptsize{\text{ Amplitude}}} }

\allowdisplaybreaks[4]

\begin{document}

\title{Unitarity Bounds and Basis Transformations in SMEFT: An Analysis of Warsaw and SILH Bases}

\author{Qing-Hong Cao}
\email{qinghongcao@pku.edu.cn}
\affiliation{School of Physics, Peking University, Beijing 100871, China}
\affiliation{Center for High Energy Physics, Peking University, Beijing 100871, China}

\author{Yandong Liu}
\email{ydliu@bnu.edu.cn}
\affiliation{Key Laboratory of Beam Technology of Ministry of Education, School of Physics and Astronomy, Beijing Normal University, Beijing, 100875, China}
\affiliation{Institute of Radiation Technology, Beijing Academy of Science and Technology, Beijing 100875, China}

\author{Shu-Run Yuan}
\email{sryuan@stu.pku.edu.cn}
\affiliation{School of Physics, Peking University, Beijing 100871, China}

\begin{abstract}
The equivalence between the Warsaw and SILH bases in Standard Model Effective Field Theory
is well established, with transformation rules connecting the two via equations of motion and field
redefinitions. This study presents an explicit calculation of the analytical unitarity bounds—defined as the marginal limit of the parameter space—for dimension-six operators within both the Warsaw and SILH bases. We employ a coupled channel analysis to scrutinize scattering processes involving vector bosons and fermions.
We conduct a comprehensive investigation into the transformation of unitarity bounds under changes in the operator basis. 
Our findings demonstrate that the transformation rules, as implicated by the equivalence theorem, can be directly applied to convert unitarity bounds from one basis to another, provided that the operators involved in the transformation rules do not belong to the same subset defined by a block in the coupled channel matrix.
\end{abstract}

\maketitle

The Standard Model Effective Field Theory (SMEFT) provides a systematic framework to describe possible effects of new physics (NP) beyond the Standard Model (SM) at energies below the NP scale $\Lambda$. By extending the SM Lagrangian with higher-dimensional operators constructed from SM fields, SMEFT allows us to parameterize NP effects in a model-independent way. These operators are suppressed by powers of $\Lambda$ and characterized by Wilson coefficients $\mathcal{C}_i$. A crucial aspect of SMEFT is the choice of operator basis. While different bases are physically equivalent, as they are related by field redefinitions and equations of motion (EoMs)~\cite{Politzer:1980me,Arzt:1993gz,Georgi:1994qn,Einhorn:2013kja,Falkowski:2015wza}, practical calculations and interpretations can vary depending on the basis used. Two commonly used bases are the Warsaw basis~\cite{Grzadkowski:2010es} and the Strongly Interacting Light Higgs (SILH) basis~\cite{Giudice:2007fh,Masso:2012eq}. Understanding how physical constraints, such as unitarity bounds, manifest in different bases is important for consistent analyses.

In this work, we perform an explicit calculation of the analytical unitarity bounds for dimension-six operators in both the Warsaw and SILH bases while previous numerical studies of unitarity bounds have been conducted in works such as~\cite{Corbett:2014ora,Corbett:2017qgl,Bilchak:1987cp,Gounaris:1993fh,Gounaris:1994cm,Gounaris:1995ed,Degrande:2013mh,Baur:1987mt,Rauch:2016pai}. By performing a coupled channel analysis for the scattering processes of vector bosons ($V=W^\pm,Z,\gamma,h$) and fermions ($f=e^-,\nu,u,d$), we derive constraints on the Wilson coefficients that ensure perturbative unitarity is preserved up to the cutoff scale $\Lambda$. Following common practice, the unitarity bounds for each Wilson coefficient are defined as the individual maximal values, obtained by marginalizing over other coefficients, as the complete parameter space is not practical for illustrating constraints.
While the full bounded parameter regions are theoretically equivalent across different parametrizations implied by the two bases, this equivalence does not necessarily translate into practical convenience in applications.
We explore the relationship between commonly used unitarity bounds in different bases and assess whether the transformation rules, derived from the equivalence of the two bases, can serve as a shortcut for transforming bounds from one basis to another, though this is generally no from simple thinking.
We discover that direct transformation of unitarity bounds is feasible when the operators involved in the relevant transformation rules belong to different subsets, as defined by the blocks in the coupled channel matrix.
Explicitly, we enumerate all subsets of the selected operator set within the unitarity bounds problem, providing possible convenience for others.

In the SMEFT, higher-dimensional operators are constructed from SM fields and organized according to their mass dimension. For dimension-six operators, there are 76 independent terms when baryon and lepton number conservation is assumed~\cite{Grzadkowski:2010es}. After accounting for Hermitian conjugation and flavor symmetries, 59 independent operators remain. The Warsaw basis~\cite{Grzadkowski:2010es} is a commonly used operator basis for dimension-six operators. It is constructed to eliminate redundancies using EoMs and to provide a minimal and complete set of operators. The SILH basis~\cite{Giudice:2007fh,Masso:2012eq}, on the other hand, is motivated by models where the Higgs boson is a pseudo-Nambu-Goldstone boson, and emphasizes operators that modify Higgs couplings. The two bases are related by EoMs and field redefinitions. The transformation between the bases involves substituting certain operators in one basis with combinations of operators in the other basis. For example, the Warsaw operator $\mathcal{O}_{\varphi W}$ is related to operators in the SILH basis via EoMs. The transformation rules can be derived by considering the equivalence of the $S$-matrix elements calculated in different bases. When performing calculations, one must ensure that physical observables are independent of the operator basis chosen.

In this study, we analytically examine the unitarity bounds of SMEFT of scattering processes of $VV\to VV$ and $f\bar{f}\to VV$ within the Warsaw basis and the SILH basis to ascertain how the equivalence principle applies in the context of unitarity analysis, where $V$ and $f$ denotes $W^\pm/Z/\gamma/h$ and $e^-/\nu/u/d$, respectively.
We perform a coupled channel analysis of scattering processes of $VV\to VV$ and $f\bar{f}\to VV$
and derive unitarity bounds on the operators.
We first adopt the standard Warsaw basis~\cite{Grzadkowski:2010es} and consider only CP-even operators listed in Table~\ref{convention_Warsaw_basis}.
Note that the four-fermion operators are not considered as they are relatively independent of other operators, meaning that cancellation is rare, and the processes of $ff\to ff$ induced by four-fermion operators are loosely related with SM electroweak sector.
In the fermion sector, the operator being generation independent is understood, e.g., $\mathcal{O}_{\varphi e}(1,1)= \mathcal{O}_{\varphi e}(2,2)= \mathcal{O}_{\varphi e}(3,3)$ and 
$\mathcal{O}_{\varphi e}(m,n)=0$ for $m\neq n$.

\begin{table}
\caption{The CP-even operators that would be constrained by unitarity  in the Warsaw basis~\cite{Grzadkowski:2010es}.}
\label{convention_Warsaw_basis}
\begin{tabular}{|c|c||c|c|}
\hline
$\mathcal{O}_W$ & $\epsilon^{IJK}W^{I\nu}_{\mu}W^{J\rho}_{\nu}W^{K\mu}_{\rho}$ & $ \mathcal{O}_{\varphi\Box}$ & $(\varphi^\dagger\varphi)\Box(\varphi^\dagger\varphi)$\\
\hline
$\mathcal{O}_{\varphi D}$ & $(\varphi^\dagger D^{\mu}\varphi)^{*}(\varphi^\dagger D_{\mu}\varphi)$ & $\mathcal{O}_{\varphi W}$ & $\varphi^\dagger\varphi W^{I}_{\mu\nu}W^{I\mu\nu}$\\
\hline
$ \mathcal{O}_{\varphi B}$ & $\varphi^\dagger\varphi B_{\mu\nu}B^{\mu\nu}$ & $ \mathcal{O}_{\varphi WB}$ & $\varphi^\dagger\tau^{I}\varphi W^{I}_{\mu\nu}B^{\mu\nu}$\\
\hline
$\mathcal{O}^{(1)}_{\varphi l}$ & $(\varphi^\dagger i \overleftrightarrow{D}_{\mu}\varphi)(\bar{l}_p \gamma^{\mu}l_r)$ & $\mathcal{O}^{(3)}_{\varphi l}$ & $(\varphi^\dagger i \overleftrightarrow{D}^{I}_{\mu}\varphi)(\bar{l}_p\tau^{I} \gamma^{\mu}l_r)$ \\
\hline 
$ \mathcal{O}_{\varphi e}$ & $(\varphi^\dagger i \overleftrightarrow{D}_{\mu}\varphi)(\bar{e}_p \gamma^{\mu}e_r)$ &$\mathcal{O}^{(1)}_{\varphi q}$ & $(\varphi^\dagger i \overleftrightarrow{D}_{\mu}\varphi)(\bar{q}_p \gamma^{\mu}q_r)$ \\
\hline
$\mathcal{O}^{(3)}_{\varphi q}$ & $(\varphi^\dagger i \overleftrightarrow{D}^{I}_{\mu}\varphi)(\bar{q}_p\tau^{I} \gamma^{\mu}q_r)$ & $ \mathcal{O}_{\varphi u}$ & $(\varphi^\dagger i \overleftrightarrow{D}_{\mu}\varphi)(\bar{u}_p \gamma^{\mu}u_r)$\\
\hline
$ \mathcal{O}_{\varphi d}$ & $(\varphi^\dagger i \overleftrightarrow{D}_{\mu}\varphi)(\bar{d}_p \gamma^{\mu}d_r)$ & $ \mathcal{O}_{\varphi ud}$ & $(\widetilde{\varphi}^\dagger i D_{\mu}\varphi)(\bar{u}_p \gamma^{\mu}d_r)$ \\
\hline
\hline
$ \mathcal{O}_{eW}$ & $(\bar{l}_{p}\sigma^{\mu\nu}e_{r})\tau^{I}\varphi W^{I}_{\mu\nu}$ & $ \mathcal{O}_{eB}$ & $(\bar{l}_{p}\sigma^{\mu\nu}e_{r})\varphi B_{\mu\nu}$ \\
\hline
$ \mathcal{O}_{uW}$ & $(\bar{q}_{p}\sigma^{\mu\nu}u_{r})\tau^{I}\widetilde{\varphi} W^{I}_{\mu\nu}$ & $ \mathcal{O}_{uB}$ & $(\bar{q}_{p}\sigma^{\mu\nu} u_{r})\widetilde{\varphi} B_{\mu\nu}$ \\
\hline
$ \mathcal{O}_{dW}$ & $(\bar{q}_{p}\sigma^{\mu\nu}d_{r})\tau^{I}\varphi W^{I}_{\mu\nu}$ & $ \mathcal{O}_{dB}$ & $(\bar{q}_{p}\sigma^{\mu\nu} d_{r})\varphi B_{\mu\nu}$\\
\hline
\end{tabular}
\end{table}

For the SILH basis~\cite{Elias-Miro:2013mua}, six operators are added in the Warsaw basis with six operators eliminated,
\begin{align}\label{trade_operator_SILH}
&(\mathcal{O}^{\prime}_{W},\mathcal{O}_{B},\mathcal{O}_{2W},\mathcal{O}_{2B},\mathcal{O}^{\rm SILH}_{W},\mathcal{O}^{\rm SILH}_{B}) \nn \\
\leftrightarrow ~ &(\mathcal{O}_{\varphi W},\mathcal{O}_{\varphi WB},\mathcal{O}_{\varphi\Box},\mathcal{O}_{\varphi D},\mathcal{O}_{\varphi l}^{(1)}, \mathcal{O}_{\varphi l}^{(3)}),
\end{align}
where 
\begin{align}
&\mathcal{O}^{\prime}_{W}\equiv \left(D_{\mu }\varphi\right){}^{\dagger } \left(i g \frac{\tau ^I}{2} W^{I;\mu  \nu }\right)\left(D_{\nu }\varphi\right),\nn\\
&\mathcal{O}_{B}\equiv  \left(D_{\mu }\varphi\right){}^{\dagger }\left(i \frac{g^{\prime}}{2} B^{\mu \nu }\right)\left(D_{\nu }\varphi\right), \nn \\
&\mathcal{O}_{2W}\equiv-\frac{1}{2}(D^\mu W^I_{\mu\nu})^2\sim \mathcal{O}_{DW},\nn\\
&\mathcal{O}_{2B}\equiv-\frac{1}{2}(\partial^\mu B_{\mu\nu})^2\sim\mathcal{O}_{DB}\nn\\
&\mathcal{O}^{\rm SILH}_{W}\equiv\frac{i g}{2}(\varphi^\dagger \tau^I \overleftrightarrow{D}^{\mu} \varphi)(D^\nu W^I_{\mu\nu}),\nn\\
&\mathcal{O}^{\rm SILH}_{B}\equiv\frac{i g^{\prime}}{2}(\varphi^\dagger \overleftrightarrow{D}^{\mu} \varphi)(\partial^\nu B_{\mu\nu}).
\end{align}
The two bases are connected by the EoMs, which between the SILH basis and the Warsaw basis read as
\begin{small}
\bea
&&2 \mathcal{O}^{\prime}_W-\frac{1}{4}g^2 \mathcal{O}_{\varphi W}-\frac{1}{4}g^{\prime } g \mathcal{O}_{\varphi WB}+\frac{3}{4}g^2 \mathcal{O}_{\varphi\Box}\nn\\
&=&-\frac{g^2}{4}\left[ \mathcal{O}^{(3)}_{\varphi l}+ \mathcal{O}^{(3)}_{\varphi q}\right] + \boxed{E},\nn \\ 
&&2 \mathcal{O}_B-\frac{1}{4}g^{\prime 2} \mathcal{O}_{\varphi B}-\frac{1}{4}g^{\prime } g \mathcal{O}_{\varphi WB}+g^{\prime  2}\left[\mathcal{O}_{\varphi D}+\frac{1}{4} \mathcal{O}_{\varphi\Box}\right]\nn\\
&=&-\frac{g^{\prime 2}}{2}\left[-\frac{1}{2} \mathcal{O}^{(1)}_{\varphi l}+\frac{1}{6} \mathcal{O}^{(1)}_{\varphi q}- \mathcal{O}_{\varphi e}+\frac{2}{3} \mathcal{O}_{\varphi u}-\frac{1}{3} \mathcal{O}_{\varphi d}\right] + \boxed{E},\nn\\ 
&&\mathcal{O}_{2W}+\frac{3g^2}{8} \mathcal{O}_{\varphi\Box} = -\frac{g^2}{4}\left[\mathcal{O}^{(3)}_{\varphi l}+\mathcal{O}^{(3)}_{\varphi q} \right] + \boxed{E},\nn\\ &&\mathcal{O}_{2B}+\frac{g^{\prime 2}}{2}\left[\frac{1}{4}\mathcal{O}_{\varphi\Box}+\mathcal{O}_{\varphi D}\right]\nn\\
&=&-\frac{g^{\prime 2}}{2}\left[-\frac{1}{2} \mathcal{O}^{(1)}_{\varphi l}+\frac{1}{6} \mathcal{O}^{(1)}_{\varphi q}- \mathcal{O}_{\varphi e}+\frac{2}{3} \mathcal{O}_{\varphi u}-\frac{1}{3} \mathcal{O}_{\varphi d}\right] + \boxed{E},\nn\\ 
&&\mathcal{O}^{\rm SILH}_W + \frac{3 g^2}{4}\mathcal{O}_{\varphi\Box} = -\frac{g^2}{4}\left[ \mathcal{O}^{(3)}_{\varphi l}+ \mathcal{O}^{(3)}_{\varphi q}\right] + \boxed{E},\nn\\ 
&&\mathcal{O}^{\rm SILH}_B + g^{\prime  2}\left[\frac{1}{4}\mathcal{O}_{\varphi\Box}+\mathcal{O}_{\varphi D}\right]\nn\\
&=& -\frac{g^{\prime 2}}{2}\left[-\frac{1}{2} \mathcal{O}^{(1)}_{\varphi l}+\frac{1}{6} \mathcal{O}^{(1)}_{\varphi q}- \mathcal{O}_{\varphi e}+\frac{2}{3} \mathcal{O}_{\varphi u}-\frac{1}{3} \mathcal{O}_{\varphi d}\right] + \boxed{E},\nn\\
\label{EoM_Eq}
\eea
\end{small}
where $\boxed{E}$ stands for those operators that have null contribution to the $VV\to VV$ or $f\bar{f}\to VV$ scattering amplitudes at the order of $O(s)$, e.g. $\mathcal{O}_{e\varphi}=(\varphi^\dagger\varphi)(\bar{l}e\varphi)$, or those four-fermion operators.

The transformation rules of helicity amplitudes between the Warsaw basis and the SILH basis are derived in Appendix \ref{App:TR}. The partial transformation rule reads as 
\begin{equation}
    \mathcal{C}_i^{\rm A}\xrightarrow[]{\scriptsize{\text{ Amplitude}}} \mathcal{C}_i^{\rm B} + \mathcal{C}_j^{\rm B},
\end{equation}
and it directly translates the helicity amplitudes in the operator basis A into those in the operator basis B. 
After decomposing the scattering helicity amplitudes into partial wave amplitudes, we derive the marginalized unitarity constraint with coupled channel analysis for each individual Wilson coefficient.

For the $V_1 V_2\to V_3 V_4$ processes, the partial wave unitarity condition~\cite{Rauch:2016pai,Baur:1987mt} for a diagonalized coupled channel matrix~\cite{Lee:1977eg} is given by
\begin{align} \label{eq:unicon1}
\abs{\text{Re}(a^J(V'_i(\ld_i)V'_{j}(\ld_{j})\to V'_i(\ld_i)V'_{j}(\ld_{j}))}\leq 1,
\end{align}
where $V'_i(\ld_i)=\sum_{a,\ld_a}y^{i,\ld_i}_{a,\ld_a}V_a(\ld_a)$ is a linear combination of states that diagonalize the field space under the same quantum number $(Q, J)$ (total charge and total angular momentum), and $a^J$ is the partial wave amplitude.
For the $f_1\bar{f_2}\to V_3V_4$ processes, we have the constraint, for each $J$,
\begin{align} \label{eq:unicon2}
\abs{a^J_{V'_j V'_k; x^i}}\leq \frac{1}{2},
\end{align}
where $x^i$ represents linearly recombined states of fermions, i.e., $\ket{x^i}=\sum_{f\sg}x^i_{f\sg}\ket{f(\sg)\bar{f}(-\sg)}$, with $x^i_{f\sg}$ being the unitary transformation matrix element ($i$ is the index in the diagonalized field space), and $\ld_a(=\pm 1,0), \sigma_a(=\pm 1)$ are the helicities of the vector bosons and fermions, respectively. This condition requires the eigenvalues of the coupled channel matrix to be less than 1/2. Note that the coupled channel matrix is not square for $f\bar{f}\to VV$ processes, so we use singular value decomposition to obtain the eigenvalues $a^J_{V'_j V'_k; x^i}$.

The unitarity bounds are obtained by diagonalizing the coupled channel matrices.
The bound on each Wilson coefficient is defined as 
\begin{align} \label{eq:defuni}
&\text{max}\left\{\mathcal{C}_j ~|~\{ \mathcal{C}_j , \mathcal{C}_k \} \in S  \right\},
\end{align}
where $S$ is a hypersurface defined through Eq.~\ref{eq:unicon1} or \ref{eq:unicon2}, in which the equality holds.
The analytical unitarity bounds on the 26 operators are presented in Table~\ref{couple_channel_results}. 
The second and third columns present the unitarity bounds obtained
in the Warsaw and SILH bases, respectively. These results are also confirmed through numerical scans using MultiNest \cite{Feroz:2007kg,Feroz:2008xx,Feroz:2013hea}.

\begin{table}[b]
\caption{
 The partial wave unitarity bounds, namely the marginalized bounds for each operator in the Warsaw basis (the second column) and the SILH basis (the third column) (i.e., $\abs{\mathcal{C}_i}\leq \mathcal{C}^{max}_{i}$ with $\mathcal{C}^{max}_{i}$ being an operator-dependent constant), with the overall scaling factor $\sqrt{s}/\Lambda=1$. 
}\label{couple_channel_results}
\begin{tabular}{c|c|c}
\hline
  \multirow{2}{*}{Operator}  & \multicolumn{2}{c}{bounds on $\mathcal{C}_i$'s}\\
\cline{2-3}
& Warsaw  &SILH\\
\hline
$\mathcal{O}_W$  & $\frac{4\pi}{3 g}$  & $\frac{4\pi}{3 g}$ \\
\hline
$\mathcal{O}_{eW}$  & $4\pi$  & $4\pi$ \\
\hline
$\mathcal{O}_{eB}$  & $4\sqrt{3}\pi$  & $4\sqrt{3}\pi$ \\
\hline
$\mathcal{O}_{uW}$  & $\frac{4\pi}{\sqrt{3}}$  & $\frac{4\pi}{\sqrt{3}}$ \\
\hline
$\mathcal{O}_{uB}$  & $4\pi$ & $4\pi$ \\
\hline
$\mathcal{O}_{dW}$  & $\frac{4\pi}{\sqrt{3}}$&  $\frac{4\pi}{\sqrt{3}}$ \\
\hline
$\mathcal{O}_{dB}$ & $4\pi$ & $4\pi$ \\
\hline
$\mathcal{O}_{\varphi ud}$  & $4\sqrt{2}\pi$& $4\sqrt{2}\pi$ \\
\hline
$\mathcal{O}_{\varphi\Box}$  & $\frac{16\pi}{3}$  & / \\
\hline
$\mathcal{O}_{\varphi D}$   &	$\frac{64\pi}{3}$  & / \\
\hline
$\mathcal{O}_{\varphi W}$  &	$\frac{4 \sqrt{6}\pi}{3}$   & / \\
\hline
$\mathcal{O}_{\varphi B}$  &	$4\sqrt{2}\pi$ &  $ 4 \sqrt{6} \pi \frac{ \sqrt{3 g^4 + g^{\prime 4}} }{3 g^2} + 8 \pi \frac{g^\prime}{ g} $ \\
\hline
$\mathcal{O}_{\varphi WB}$  & $8\pi$   & / \\
\hline
$\mathcal{O}^{\prime}_{W}$  & /  & $\frac{32\sqrt{6}\pi}{3 g^2}$ \\
\hline
$\mathcal{O}_{B}$  & /  & $ \frac{64 \pi}{g g^\prime} + \frac{32 \sqrt{6}\pi }{3g^2} $ \\
\hline
$\mathcal{O}_{2W}$  & /  & $\frac{16(4+3\sqrt{6})\pi}{3g^{ 2}} $ \\
\hline
$\mathcal{O}_{2B}$ & /  & $\frac{16(8+3\sqrt{6})\pi}{3g^{\prime 2}} $ \\
\hline
$\mathcal{O}^\text{SILH}_{W}$ & / & $\frac{8(8+3\sqrt{6})}{3g^{ 2}} + \frac{64\pi}{11g^2}$ \\
\hline
$\displaystyle \mathcal{O}^\text{SILH}_{B}$ & / &  $\frac{8(16+3\sqrt{6})}{3g^{\prime 2}} +\frac{8\pi}{g^2} + \frac{24\pi g^{\prime 6}}{ g^4 (8g^4 + 3g^{\prime 4} ) } $\\
\hline
$\mathcal{O}^{(1)}_{\varphi l}$ &$2\sqrt{6}\pi$ & / \\
\hline
$\mathcal{O}^{(3)}_{\varphi l}$  & $2\sqrt{6}\pi$   & / \\
\hline
$\mathcal{O}_{\varphi e}$  & $4\sqrt{3}\pi$ &  $12\pi$ \\
\hline
$\mathcal{O}^{(1)}_{\varphi q}$  & $2\sqrt{2}\pi$ &  $4\sqrt{\frac{2}{3}}\pi$ \\
\hline
$\mathcal{O}^{(3)}_{\varphi q}$  & $2\sqrt{2}\pi$ &  $4\sqrt{2}\pi$ \\
\hline
$\mathcal{O}_{\varphi u}$  & $4\pi$ &  $4\sqrt{\frac{11}{3}}\pi$ \\
\hline
$\mathcal{O}_{\varphi d}$ & $4\pi$& $4\sqrt{\frac{5}{3}}\pi$ \\
\hline
\end{tabular}
\end{table}

With the analytical unitarity bounds established, we revisit the transformation rules in Appendix \ref{App:TR} to examine their application in this context. The second partial transformation rules applying unitarity bounds read as
\begin{equation}
\mathcal{C}_i^{\rm A}\xleftarrow[\scriptsize{\text{Constraint}}]{} \mathcal{C}_i^{\rm B} + \mathcal{C}_j^{\rm B},
\end{equation}
which translate the unitarity bounds in the operator basis B into the bounds in the operator basis A. However, it is unclear whether the aforementioned methodology would yield identical bounds for operators in the basis A from corresponding operators in the basis B to those derived directly. Next, we will probe into the applicability of these transformation rules in determining unitary bounds for one basis from another.

\begin{figure}
    \centering
    \includegraphics[scale=0.4]{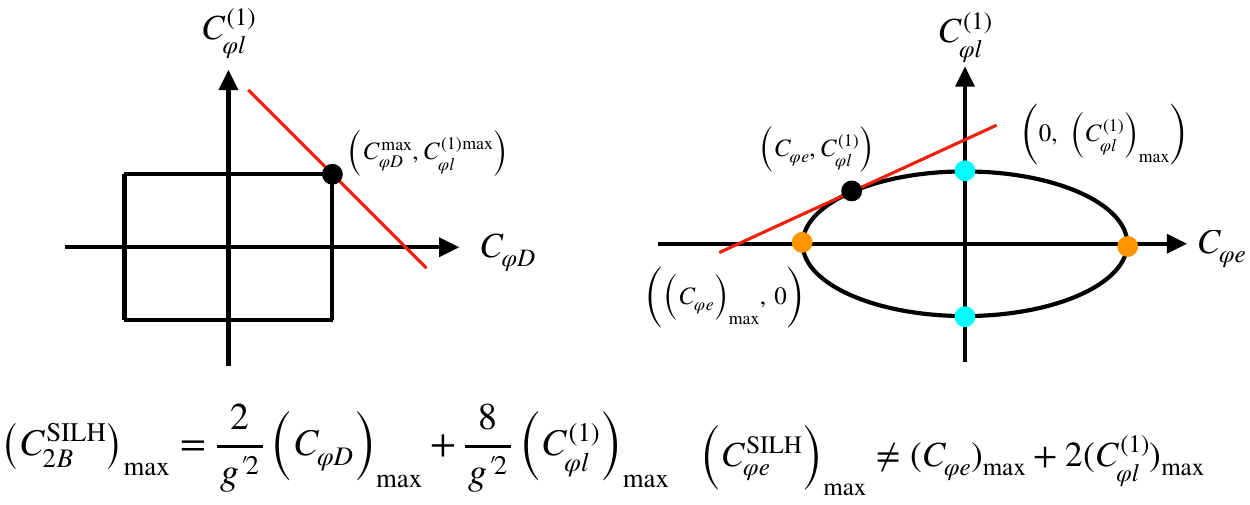}
    \caption{Schematic diagram of unitary bound on operators belong to the different and same sets in the Warsaw basis.}
    \label{fig:operatorset}
\end{figure}

We begin with deriving unitarity bounds in the SILH basis from those in the Warsaw basis with the help of transformation rules. We find that the transformation rules successfully reproduce the correct unitarity bounds in the SILH basis if all the operators involved in the transformation rules can reach their unitarity bounds simultaneously.
For example, consider the operator $\mathcal{O}_{2B}$, which appears only in the SILH basis and is related to $\mathcal{O}_{\varphi l}^{(1)} $ and $\mathcal{O}_{\varphi D}$ in the Warsaw basis as following: 
\begin{equation}
\mathcal{C}_{2B} =\frac{8}{g^{\prime 2}} \mathcal{C}_{\varphi l}^{(1)} +\frac{2}{g^{\prime 2}} \mathcal{C}_{\varphi D}.
\label{eq:o2b_tranrule}
\end{equation}
Substituting the bounds of $\mathcal{O}_{\varphi l}^{(1)} $ and $\mathcal{O}_{\varphi D}$ in the Warsaw basis,
\begin{equation}
(\mathcal{C}_{\varphi l}^{(1)})_{\rm max} =2\sqrt{6}\pi,\qquad (\mathcal{C}_{\varphi D})_{\rm max}=\frac{64\pi}{3},
\end{equation}
into Eq.~\ref{eq:o2b_tranrule}, we obtain the bound $\left(\mathcal{C}_{2B}\right)_{\rm max}$ in the SILH basis,
\begin{equation*}
\left(\mathcal{C}_{2B}\right)_{\rm max}=\frac{16 (8+3\sqrt{6}) \pi}{3 g^{\prime 2} }.
\end{equation*}
The left panel of Fig.~\ref{fig:operatorset} shows the unitarity bounds of $\mathcal{O}_{\varphi l}^{(1)} $ and $\mathcal{O}_{\varphi D}$ in the Warsaw basis; see the rectangle region surrounded by the black lines. The red line denotes the transformation rule given in Eq.~\ref{eq:o2b_tranrule}, and it exhibits a maximal intercept when crossing the black point, where both the $\mathcal{C}_{\varphi l}^{(1)} $ and $\mathcal{C}_{\varphi D}$ reach their unitarity bounds simultaneously.

Unfortunately, the unitarity bounds of a few operators in the SILH basis cannot be derived from the transformation rule. For instance, the operator $\mathcal{O}_{\varphi e}$ in the SILH basis is related to 
$\mathcal{O}_{\varphi e}$ and $\mathcal{O}_{\varphi l}^{(1)}$ in the Warsaw basis, i.e.,
\begin{equation}
\mathcal{C}_{\varphi e}^{\rm SILH} = \mathcal{C}_{\varphi e}^{\rm Warsaw} -2 \mathcal{C}_{\varphi l}^{(1){\rm Warsaw}}.
\label{eq:tranrule2}
\end{equation}
Substituting the unitarity bounds $(\mathcal{C}_{\varphi e})_{\rm max}=4\sqrt{3}\pi$ and $(\mathcal{C}_{\varphi l}^{(1)})_{\rm max}=2\sqrt{6}\pi$ in the Warsaw basis into the above equation does not generate the correct $(\mathcal{C}_{\varphi e})_{\rm max}$ in the SILH basis. It is due to the fact that the two operators are strongly correlated in the coupled channel analysis such that the two operators in the Warsaw basis never reach their unitary bounds at the same time. 
The ellipse in the right panel of Fig.~\ref{fig:operatorset} represents the unitarity bounds on $\mathcal{O}_{\varphi e}$ and $\mathcal{O}_{\varphi l}^{(1)}$ in the Warsaw basis, i.e.,
\begin{equation}
\left(\mathcal{C}_{\varphi e}\right)^2 +2 \left(\mathcal{C}_{\varphi l}^{(1)}\right)^2=48\pi^2.
\end{equation}
The transformation rule in Eq.~\ref{eq:tranrule2} is denoted by the red line, and the unitarity bounds on $\mathcal{C}_{\varphi e}$ in the SILH basis is the maximum intercept of red line. It occurs when the red line is tangent to the ellipse, and the tangency point (TP) is ($\mathcal{C}_{\varphi e}$,~$\mathcal{C}_{\varphi l}^{(1)}$)=($4\pi$,~$-4\pi$). Substituting the tangency point into the transformation rule in Eq.~\ref{eq:tranrule2},  we obtain 
\begin{align}
(\mathcal{C}_{\varphi e})_{\rm max}=(\mathcal{C}_{\varphi e})_{\rm TP} -2 (\mathcal{C}_{\varphi l}^{(1)})_{\rm TP}=12\pi,
\end{align}
which is the correct $(\mathcal{C}_{\varphi e})_{\rm max}$ in the SILH basis.

Whether the operator in the transformation rule can reach each individual unitarity bound depend on their roles in the coupled channel analysis.
The coupled channel matrix is block diagonalized to obtain the unitarity bounds on operators, and each block itself respects unitarity. We classify the 20 operators into six sets based on their appearance in the block. 
All of the operators in the same set must be taken into account to obtain the correct bounds, as interference effects prohibit operators from reaching their unitarity bounds simultaneously. However, operators from different sets can always reach their unitarity bounds at the same time, as the operators are decomposed into independent and orthogonal subspaces in the coupled channel analysis. 
In the Warsaw basis, the operator sets are
\begin{align} \label{eq:warsawsets}
&\{ {\cal{O}}_{W}, \mathcal{O}_{\varphi \Box}, \mathcal{O}_{\varphi W}, \mathcal{O}_{\varphi B} \}, \nn \\
&\{ \mathcal{O}_{\varphi D}, \mathcal{O}_{\varphi \Box},  \mathcal{O}_{\varphi WB} \}, \nn \\
&\{ {\cal{O}}_{W}, {\cal{O}}_{\varphi l}^{(3)}, {\cal{O}}_{\varphi q}^{(3)}   \}, \nn \\
&\{ {\cal{O}}_{\varphi d},  {\cal{O}}_{\varphi u},  {\cal{O}}_{\varphi l}^{(1)},  {\cal{O}}_{\varphi q}^{(1)}, {\cal{O}}_{\varphi e}        \}, \nn \\
&\{{\cal{O}}_{uW}, {\cal{O}}_{uB}, {\cal{O}}_{dW}, {\cal{O}}_{dB}, {\cal{O}}_{eW}, {\cal{O}}_{eB}    \}, \nn \\
&\{{\cal{O}}_{\varphi ud}  \}, \nn\\
&\{{\cal{O}}_{W} \},
\end{align}
and in the SILH basis
\begin{align} \label{eq:silhsets}
&\{ {\cal{O}}_{W}, \mathcal{O}_{\varphi B}, \mathcal{O}^{\prime}_{W}, \mathcal{O}_{B}, \mathcal{O}_{2W}, \mathcal{O}_{2B}, \mathcal{O}^\text{SILH}_{W}, \mathcal{O}^\text{SILH}_{B}  \}, \nn \\
&\{  \mathcal{O}^{\prime}_{W}, \mathcal{O}_{B}, \mathcal{O}_{2W}, \mathcal{O}_{2B}, \mathcal{O}^\text{SILH}_{W}, \mathcal{O}^\text{SILH}_{B}  \}, \nn \\
&\{ {\cal{O}}_{W}, {\cal{O}}_{\varphi q}^{(3)}, \mathcal{O}^{\prime}_{W}, \mathcal{O}^\text{SILH}_{W}, \mathcal{O}_{2W} \}, \nn \\
&\{ {\cal{O}}_{\varphi d},  {\cal{O}}_{\varphi u},   {\cal{O}}_{\varphi q}^{(1)}, {\cal{O}}_{\varphi e}, \mathcal{O}_{B}, \mathcal{O}^\text{SILH}_{B}\}, \nn \\
&\{{\cal{O}}_{uW}, {\cal{O}}_{uB}, {\cal{O}}_{dW}, {\cal{O}}_{dB}, {\cal{O}}_{eW}, {\cal{O}}_{eB}    \}, \nn \\
&\{{\cal{O}}_{\varphi ud}  \},\nn\\
&\{{\cal{O}}_{W} \}.
\end{align}

In this work, we analyzed the unitarity bounds in the Warsaw and SILH bases for the $ff \to VV$ and $VV \to VV$ scattering processes.
We explicitly calculated the marginal limit of each Wilson coefficient within the bounded parameter space and explored the transformation of these bounds across different bases.
Specifically, we found that in cases where all operators involved in the transformation rule can simultaneously reach their unitarity bounds, the transformation rules allow for a direct conversion of these bounds between the bases. Conversely, in situations where the operators cannot reach their bounds simultaneously, it becomes necessary to determine an optimal point on the boundary of the unitarity bound space to reproduce the correct bounds in the other basis. The two cases classification depends on whether the operators considered belong to the same subset, as derived in Eqs.~\ref{eq:warsawsets} and~\ref{eq:silhsets}. This highlights both the utility and the limitations of using transformation rules to interpret unitarity bounds across different operator bases in SMEFT.

\noindent{\it Acknowledgements}: We thank Jue Zhang and Ya Zhang for involvement in the early stage of this work. We thank Jiang-Hao Yu and Bin Yan for the valuable comments. The work is partly supported by the National Science Foundation of China under Grant Nos. 11635001, 11675002, 11725520, 11805013, 12075257, and 12235001.

\appendix
\section{Transformation rules of operator bases\label{App:TR}}

It is shown that the equivalence theorem ensures the $S$-matrix calculated in various operator bases can be mutually transformed through the Equations of Motion (EoMs) or field redefinitions~\cite{Politzer:1980me,Arzt:1993gz,Georgi:1994qn,Einhorn:2013kja}.
Below, we derive the transformation rules for the Wilson coefficients between the two operator bases,  and the rules are process independent. The derivation is grounded on the completeness and independence of the operator basis.
Consider two operator sets $\{\mathcal{O}^A_1, \mathcal{O}_j, \mathcal{O}^{\rm{EoM}}_j\}$ and $\{\mathcal{O}^B_1, \mathcal{O}_j, \mathcal{O}^{\rm{EoM}}_j\}$ connected through the EoMs. Here the superscript EoM denotes the operators appearing in the EoMs. Without loss of generality, we consider only one operator replaced between two sets, namely $\mathcal{O}^A_1\leftrightarrow \mathcal{O}^B_1$, and the EoM reads as 
\begin{align} \label{EQ:EOMAB}
    a_1 \mathcal{O}^A_1 + b_1 \mathcal{O}^B_1 + \sum_j c_j  \mathcal{O}^{\rm{EoM}}_j=0.
\end{align}
For a process of $|i\rangle\to |f\rangle$, the helicity amplitude is  
\begin{align}\label{EQ:MA}
    \mathcal{M}=\mathcal{C}^A_1 \mathcal{M}^A_1 + \sum_j \mathcal{C}_j \mathcal{M}_j + \sum_j \mathcal{C}^{\rm{EoM}}_j \mathcal{M}^{\rm{EoM}}_j
\end{align}
in one operator basis A, where $\mathcal{M}_j\equiv \langle f| O_j | i \rangle$, and 
\begin{align}\label{EQ:MB}
    \mathcal{M}=\mathcal{C}^B_1 \mathcal{M}^B_1 + \sum_j \mathcal{C}_j \mathcal{M}_j + \sum_j \mathcal{C}^{\rm{EoM}}_j \mathcal{M}^{\rm{EoM}}_j
\end{align}
in another basis B. The EoM reads as 
\begin{align} \label{EQ:MEOM}
     a_1 \mathcal{M}^A_1 + b_1 \mathcal{M}^B_1 + \sum_j c_j  \mathcal{M}^{\rm{EoM}}_j=0.
\end{align}
Utilizing the amplitude relation from the EoM yields 
\begin{align} 
    \mathcal{M}_1^B = -\frac{a_1}{b_1}\mathcal{M}^A_1 -\sum_j \frac{c_j}{b_1} \mathcal{M}^{\rm{EoM}}_j,
\end{align}
then the helicity amplitude $\mathcal{M}$ in the basis B becomes 
\begin{align} \label{EQ:MA2}
    \mathcal{M}= -\frac{a_1}{b_1}C^B_1 \mathcal{M}_1^A+ \sum_j \mathcal{C}_j \mathcal{M}_j + \sum_j (\mathcal{C}^{\rm{EoM}}_j-\frac{c_j}{b_1}\mathcal{C}^B_1 ) \mathcal{M}^{\rm{EoM}}_j.
\end{align}
As a result of the equivalence of the operator bases, Eqs.~\ref{EQ:MA} and~\ref{EQ:MA2} must be the same, and it yields the transformation rules 
\begin{align} \label{eq:gentran}
    &\mathcal{C}_1^A \to -\frac{a_1}{b_1}\mathcal{C}^B_1, \nonumber \\
    &\mathcal{C}^{\rm{EoM}}_j \to \mathcal{C}^{\rm{EoM}}_j-\frac{c_j}{b_1}\mathcal{C}^B_1,
\end{align}
which transform the helicity amplitude in basis A to that in basis B.

The transformation rules of helicity amplitudes are shown below, and the transformation rules of the operator constraints go in the opposite direction. The rules from the Warsaw basis to the SILH basis read as
\begin{small}
\begin{align}
\mathcal{C}_{\varphi W} & \lr \frac{g^2}{8}\mathcal{C}^{\prime}_{W},\nn\\
\mathcal{C}_{\varphi B} &\lr \mathcal{C}_{\varphi B} +\frac{g^{\prime 2}}{8}\mathcal{C}_{B},\nn\\
\mathcal{C}_{\varphi WB} &\lr  \frac{g g^\prime}{8}( \mathcal{C}^{\prime}_{W} +\mathcal{C}_{B} ),\nn\\
\mathcal{C}_{\varphi D} &\lr - g^{\prime 2} \mathcal{C}^\text{SILH}_{B} -\frac{g^{\prime 2}  }{2}(\mathcal{C}_{B} + \mathcal{C}_{2B} ),\nn\\
\mathcal{C}^{(1)}_{\varphi l} &\lr \frac{g^{\prime 2}  }{8 } \mathcal{C}_{B}+ \frac{ g^{\prime 2}  }{4}(\mathcal{C}^\text{SILH}_{B} + \mathcal{C}_{2B} ) ,\nn\\
\mathcal{C}^{(3)}_{\varphi l} &\lr  - \frac{g^{ 2}  }{8 } \mathcal{C}^{\prime}_{W}- \frac{g^2  }{ 4} ( \mathcal{C}^\text{SILH}_{W } + \mathcal{C}_{2W}  ),\nn\\ 
 \mathcal{C}^{(1)}_{\varphi q} &\lr  \mathcal{C}^{(1)}_{\varphi q} -\frac{g^{\prime 2}}{24}\mathcal{C}_{B}  -
\frac{g^{\prime 2} }{12}(\mathcal{C}^\text{SILH}_{B} + \mathcal{C}_{2B}),\nn\\ 
 \mathcal{C}^{(3)}_{\varphi q} & \lr  \mathcal{C}^{(3)}_{\varphi q} -\frac{g^{ 2}}{8}\mathcal{C}^{\prime}_{W}  - \frac{g^{ 2} }{4}(\mathcal{C}^\text{SILH}_{W} + \mathcal{C}_{2W}),\nn\\
\mathcal{C}_{\varphi\Box} &\lr - \frac{3 g^2  }{4 }\mathcal{C}^\text{SILH}_{W } -\frac{3 g^2 }{8}( \mathcal{C}^{\prime}_{W } + \mathcal{C}_{2W} )\nn\\
&~\quad - \frac{ g^{\prime 2}  }{4 }\mathcal{C}^\text{SILH}_{B } -\frac{ g^{\prime 2} }{8}( \mathcal{C}_{B } + \mathcal{C}_{2B} ),\nn\\  
\mathcal{C}_{\varphi e} & \lr \mathcal{C}_{\varphi e} +\frac{g^{\prime 2} }{4}(\mathcal{C}_{B} +2 \mathcal{C}_{2B} + 2\mathcal{C}^\text{SILH}_{B}) ,\nn\\
\mathcal{C}_{\varphi d} &\lr \mathcal{C}_{\varphi d} +\frac{g^{\prime 2} }{12}(\mathcal{C}_{B} +2 \mathcal{C}_{2B} + 2\mathcal{C}^\text{SILH}_{B}),\nn\\  
\mathcal{C}_{\varphi u} &\lr \mathcal{C}_{\varphi u} -\frac{g^{\prime 2} }{6}(\mathcal{C}_{B} +2 \mathcal{C}_{2B} + 2\mathcal{C}^\text{SILH}_{B}).
\label{eq:warsaw2SILH}
\end{align}
\end{small}
The rules from the SILH basis to the Warsaw basis are
\begin{small}
\begin{align}
\mathcal{C}^{\prime}_{W} &\lr   \frac{8}{g^2}\mathcal{C}_{\varphi W}, \nn\\
\mathcal{C}_{\varphi e} &\lr \mathcal{C}_{\varphi e} -2\mathcal{C}^{(1)}_{\varphi l} ,\nn\\
\mathcal{C}_{\varphi d} &\lr \mathcal{C}_{\varphi d } -\frac{2}{3}\mathcal{C}^{(1)}_{\varphi l}   , \nn\\
\mathcal{C}_{\varphi u } &\lr \mathcal{C}_{\varphi u } + \frac{4 }{3}\mathcal{C}^{(1)}_{\varphi l},\nn\\
\mathcal{C}^{(1)}_{\varphi q} &\lr \mathcal{C}^{(1)}_{\varphi q} +\frac{1}{3}\mathcal{C}^{(1)}_{\varphi l} , \nn\\
\mathcal{C}^{(3)}_{\varphi q} &\lr \mathcal{C}^{(3)}_{\varphi q} - \mathcal{C}^{(3)}_{\varphi l}, \nn\\
\mathcal{C}_{B}&\lr  \frac{8}{g g^\prime}\mathcal{C}_{\varphi WB} - \frac{8}{g^2}\mathcal{C}_{\varphi W}\nn \\
\mathcal{C}_{2B}&\lr \frac{8}{g^{\prime 2}}\mathcal{C}^{(1)}_{\varphi l} + \frac{2}{g^{\prime 2}}\mathcal{C}_{\varphi D} \nn \\
\mathcal{C}_{\varphi B} &\lr \mathcal{C}_{\varphi B}+\frac{g^{\prime 2}}{g^2}\mathcal{C}_{\varphi W}-\frac{g^{\prime }}{g}\mathcal{C}_{\varphi WB}\nn \\
\mathcal{C}_{2W}&\lr \frac{8}{3g^{ 2}}\mathcal{C}_{\varphi\Box} - \frac{2}{3g^{ 2}}\mathcal{C}_{\varphi D} - \frac{8}{g^{ 2}}\mathcal{C}^{(3)}_{\varphi l} \nn \\
\mathcal{C}^\text{SILH}_{B} &\lr \frac{4}{g^{ 2}}\mathcal{C}_{\varphi W} -\frac{4}{gg^\prime}\mathcal{C}_{\varphi WB}-\frac{2}{g^{\prime 2}}(\mathcal{C}_{\varphi D} +2\mathcal{C}^{(1)}_{\varphi l} ) \nn \\
\mathcal{C}^\text{SILH}_{W} &\lr \frac{2}{3g^{ 2}}(\mathcal{C}_{\varphi D} +6\mathcal{C}^{(3)}_{\varphi l}-6\mathcal{C}_{\varphi W}-4\mathcal{C}_{\varphi\Box} ).
\label{eq:SILH2warsaw}
\end{align}
\end{small}

% ================================================================================================
\bibliographystyle{apsrev}
\bibliography{reference}

\end{document}